\documentclass[10pt]{iopart}
\usepackage{iopams}  
\usepackage{epsf,amssymb,graphicx,sidecap,multirow}
\usepackage{amsfonts}
\usepackage{verbatim,color}
\begin{document}
\title{Majority-vote dynamics on multiplex networks}
\author{Jeehye~Choi and K.-I.~Goh}
\address{Department of Physics, Korea University, Seoul 02841, Korea}
\ead{kgoh@korea.ac.kr}
\vspace{10pt}
\begin{indented}
\item[] Dated: 15 November 2018
\end{indented}

\begin{abstract}
Majority-vote model is a much-studied model for social opinion dynamics of two competing opinions. With the recent appreciation that our social network comprises a variety of different ``layers" forming a multiplex network, a natural question arises on how such multiplex interactions affect the social opinion dynamics and consensus formation. Here, the majority-vote model will be studied on multiplex networks to understand the effect of multiplexity on opinion dynamics. We will discuss how global consensus is reached by different types of voters: \textit{AND}- and \textit{OR}-rule voters on multiplex-network and voters on single-network system. The \textit{AND}-model reaches the largest consensus below the critical noise parameter $q_c$. It needs, however, much longer time to reach consensus than other models. In the vicinity of the transition point, the consensus collapses abruptly. The \textit{OR}-model attains smaller level of consensus than the \textit{AND}-rule but reaches the consensus more quickly. Its consensus transition is continuous. The numerical simulation results are supported qualitatively by analytical calculations based on the approximate master equation.
\end{abstract}
\vspace{2pc}


\submitto{\NJP}

\section{Introduction}
From the network study of cascading failures on interdependent systems \cite{Buldyrev2010} to networks of networks \cite{NetONetsBook}, multiplex network \cite{Mason2014,Lee2015} has been a general theme to describe real complex systems recently.
Multiplex system is not a simply aggregated or combined system of many single networks but a functionally integrated system of them. 
Multiplex systems have shown to exhibit exotic phase transitions exhibiting, {\it e.g.}, discontinuity.   
Percolations \cite{Son2012,Lee2012NJP,Weak,Min2014,Bianconi2015}, spreading of epidemics and information~\cite{Granell2013,Min2016}, and many other standard models of interdisciplinary studies~\cite{Gomez2013,Brummitt12,Um2011,Sznaj-Weron2015} have been studied on multiplex networks \cite{Lee2015}.
Statistical-mechanics models such as ferromagnetic spin models have also been studied on multiplex and connected network setting \cite{Suchecki2009,Jang2015}. 

Social network is a prime example multiplex networks where more than one {layers} of social interactions play roles towards the societal function. Multiplex network framework is therefore essential to better understanding of structure and dynamics on social networks. 
There has been a vast body of studies using statistical physics models such as the Ising model and the voter model as a tool for addressing and solving problems in social dynamics \cite{RMPSD2009}. Following this line, here we study the problem of social consensus dynamics focused on the effect of network multiplexity. 
To this end, we use the majority-vote model \cite{Liggett1985,Oliveira1992} as a model of social consensus dynamics. Majority-vote model is a simple toy model of social consensus dynamics that could be mapped onto a non-equilibrium spin dynamics \cite{Oliveira1992}. It has been studied steadily in the complex networks literature \cite{Moreira2003,Pereira2005,Chen2017}, yet the effect of network multiplexity remains to be understood, which is the main aim of this study.

This paper is organized as follows. We give a very brief introduction to the majority-vote model in Sec.~2 and introduce the multiplex majority-vote processes studied in this work in Sec.~3. The consensus transitions of the multiplex models are studied on random regular networks in Sec.~4 and on Erd\H{o}s-R\'enyi networks in Sec.~5. Summary and discussion follow in Sec.~6. The Appendix contains the details of the approximate master equation calculations.

\section{Majority-vote process}
Many human social behaviors can be modeled by a sequence of decision processes among two alternative choices (to purchase A or B, to select A or B, and to vote for A or B, {\it etc}), which can conveniently be modeled as a binary spin state $\sigma_i=\{-1,+1\}$. 
The majority-vote model posits that a node (voter) tends to follow the majority state of its neighbors. More precisely, the system's state $\{\sigma_1,\dots,\sigma_N\}$ ($N$ denotes the number of nodes) evolves by the following dynamic rules at each step, following Ref.~\cite{Oliveira1992}: 
\begin{itemize}
\item[$i)$] A site is chosen randomly and this site looks at its nearest neighbor sites' states.
\item[$ii)$] If there is a majority-vote state taken by more than half of its neighbors, the site takes the majority-vote state with probability $1-q$, while with the rest probability $q$ it takes the opposite (minority) state. 
\item[$iii)$] In case of no majority, it takes any of the binary states with equal probability.
\end{itemize} 
Here $q$ is called the noise parameter and takes a value in the range $0\le q<1/2$. Nonzero $q$ allows fluctuations around local consensus, thus playing a role of ``temperature'' in the consensus formation dynamics. 
This decision rule can be reformulated in terms of the ``spin flip'' probability $w_i$ that the site $i$ with the current spin state $\sigma_i$ will flip its spin state at the step, which can be expressed as
\begin{equation}
w_i(\{\sigma\})=\frac{1}{2}\Big[1-(1-2q)\sigma_i \mathrm{sgn}\Big(\sum_{j\in\partial} \sigma_{j\in\partial} \Big) \Big] ,
\label{eq:w_i}
\end{equation}
where $\mathrm{sgn}(x)$ is signum function [defined as $\mathrm{sgn}(x)=1$ for $x>0$, $-1$ for $x<0$, and $0$ for $x=0$] and the summation runs over the nearest neighbors of $i$ denoted as $\partial$. 

\section{Multiplex majority-vote processes: \textit{AND}- and \textit{OR}-models}
Our aim in this paper is to investigate the effects of multiplexity to consensus formation dynamics using the majority-vote model. To this end, we generalize the majority-vote model to voters on multiplex networks. 
Let us suppose that a person interacts via two social network layers such as the family layer and the fellow workers layer.
In general the ``local'' majority opinion within each of the two layers may or may not be the same. 
Facing such multiplex social environment, the person has to make decision on which majority she would follow. 
The basic rationale of multiplex network approach is that the decision rule is formulated {\it layerwise}. 
To implement this rationale concretely, we define two multiplex decision rules, the so-called \textit{AND}- and \textit{OR}-models, for the majority-vote processes in multiplex networks.

\subsection{\textit{AND}-model}
We suppose that the \textit{AND}-voters takes the layerwise majority opinions \textit{conjunctively}, that is, they tend to follow the common majority of all the layers.
Specifically, we define the following dynamic rules at each step for the \textit{AND}-model: 
\begin{itemize}
\item[$i)$] A site is chosen randomly and this site looks at its nearest neighbor sites' states separately in each layer. 
\item[$ii)$] If there is a common majority state among all the layers, the site takes this common majority-vote state with probability $1-q$, while with the rest probability $q$ it takes the opposite (minority) state. 
\item[$iii)$] In case of no majority, it remains at its current state.
\end{itemize} 

Note the third step, where we assume that the \textit{AND}-voters do not care to update their state when there is no common majority over all layers. This is a quite strict way of applying the conjunctive rule in the decision process, which we take deliberately to discern the effect of conjunctive multiplexity to the greatest.

The spin flip probability of the \textit{AND}-model can be expressed as
\begin{eqnarray}
w_i^{\mathit{AND}}(\{\sigma\})=\frac{1}{2}&\left[1- (1-2q)\sigma_i \mathrm{sgn}\Big(\sum_{j\in\partial_1} \sigma_{j\in\partial_1} \Big) \right]\times\nonumber\\
&\times \delta\Bigg({\cal L}-\Big|\sum_{\alpha}\mathrm{sgn}\Big(\sum_{j\in\partial_{\alpha}}\sigma_{j\in\partial_{\alpha}}\Big)\Big|\Bigg)~,
\label{eq:w_i2}
\end{eqnarray}
where ${\cal L}$ is the total number of layers; $\alpha$ denotes the network layer index; $\partial_{\alpha}$ denotes the set of nearest neighbors of $i$ in the $\alpha$-layer; $\delta(x)$ denotes the delta function.

\subsection{\textit{OR}-model}

The \textit{OR}-voters takes the layerwise majority opinions \textit{disjunctively}, that is, they tend to follow the majority of any one of the layers.
Specifically, we define the following dynamic rules at each step for the \textit{OR}-model: 
\begin{itemize}
\item[$i)$] A site is chosen randomly and this site looks at its nearest neighbor sites' states in the randomly-chosen layer. 
\item[$ii)$] If there is a majority state in the chosen layer, the site takes the majority-vote state with probability $1-q$, while with the rest probability $q$ it takes the opposite (minority) state. 
\item[$iii)$] In case of no majority state in the chosen layer, it takes any of the binary states with equal probability.\end{itemize} 
The \textit{OR}-voters behave like the usual majority-voters except that they randomly switch their opinion-consulting layer at each step. 

Spin flip probability of the \textit{OR}-model can be expressed as
\begin{eqnarray}
w_i^{\mathit{OR}}(\{\sigma\})&=\frac{1}{{\cal L}} \sum_{\alpha}  \frac{1}{2}\left[1- (1-2q)\sigma_{i} \mathrm{sgn}\Big(\sum_{j\in\partial_{\alpha}} \sigma_{j\in\partial_{\alpha}} \Big) \right] \nonumber\\
&=\frac{1}{2}\left[1- (1-2q)\sigma_{i} \frac{1}{{\cal L}} \sum_{\alpha}   \mathrm{sgn}\Big(\sum_{j\in\partial_{\alpha}} \sigma_{j\in\partial_{\alpha}} \Big) \right]~.
\end{eqnarray}
From Eq.~(3), the \textit{OR}-voters may also be interpreted as the voters following the {\it average} majority opinions among all the layers.

\subsection{Numerical simulations}
We perform Monte Carlo simulations of the multiplex models on two-layer multiplex networks. 
Nodes are assigned their initial states at random. Each step, a node, say $i$, is chosen at random and the spin-flip probability $w_i$ is calculated. The node flips its spin state $\sigma_i$ stochastically with probability $w_i$. $N$ consecutive steps defines one Monte Carlo time. 
Simulation continues for a time long enough to reach the stationary state.
Typically, our simulations run for the Monte Carlo time upto $t = 2^{20}$ (AND) and $t = 2^{16}$ (OR) on networks with sizes upto $N \leq 2\times10^5$.

Our main quantity of interest is the consensus level or the ``magnetization'' $M$ of the system. We first define the instantaneous magnetization $m_t$ of the spin configuration $\{\sigma_i(t)\}$ at time $t$ as 
\begin{eqnarray}
m_t=\left|\frac{1}{N}\sum_{i=1}^N\sigma_i(t)\right|~,
\end{eqnarray}  
where $N$ is the total number of nodes. Magnetization $M$ is given by its average as 
\begin{eqnarray}
\label{eq:M}
M = \Big\langle \overline{m_t} \Big\rangle ~,
\end{eqnarray}
where $\overline{\,\cdots\,}$ denotes the temporal average calculated in the stationary state for the interval $(t_{\mathrm{final}}/2, t_{\mathrm{final}})$ where $t_{\mathrm{final}}$ is the final time of the simulation; and  $\langle \cdots \rangle$ is the ensemble average over different simulation runs and network configurations. $M$ takes the value $1$ if all the nodes are in the same spin state, that is, in complete consensus, and $0$ if the nodes are split into two equal-size groups of opposite opinions without a clear majority opinion. In that sense, $M$ may be called colloquially the consensus level of the system. 

In this study we examined the two cases of duplex networks: One with layers of random regular networks and the other with layers of Erd\H{o}s-R\'enyi networks.

\subsection{Approximate master equation calculations}
The numerical results from Monte Carlo simulations are corroborated by the analytic calculation results using the approximate master equation formalism \cite{Gleeson2013} applied to our models. The details are presented as Appendix.

\begin{figure}
\centering
\includegraphics[width=.99\linewidth]{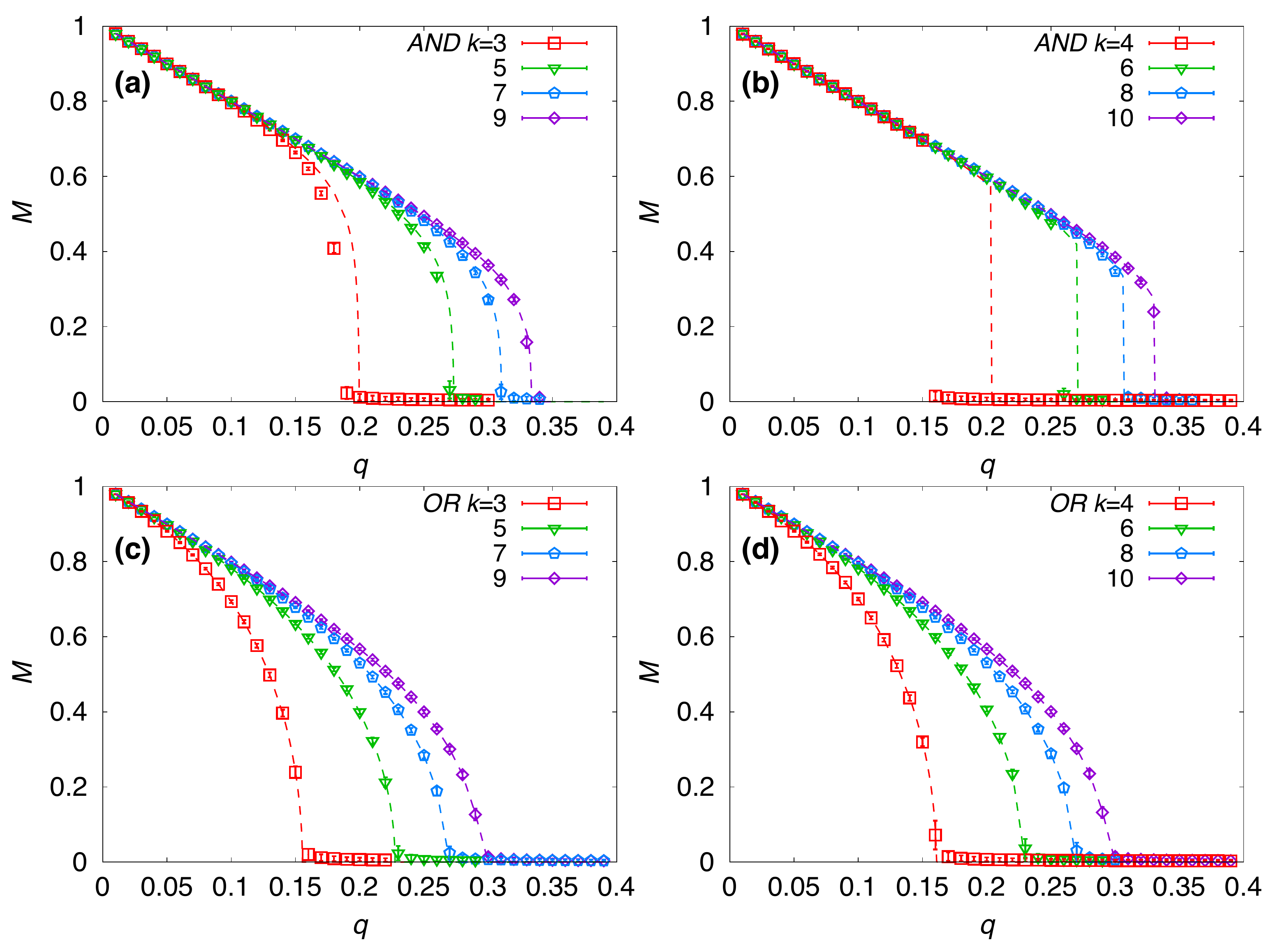}
\caption{The magnetization $M$ on $k$-RR duplexes with odd degrees $k=\{3,5,7,9\}$ and even degrees $k=\{4,6,8,10\}$.
(a) and  (b) are the results of \textit{AND}-model and (c) and (d) are those of \textit{OR}-model.
$M$ is calculated from $50$ network configurations for the Monte Carlo time $t_{\mathrm{final}}=2^{16}$ [(a), (c), (d)] and $t_{\mathrm{final}}=2^{20}$~[(b)].
Number of nodes in the network is $N=10^5$.
Points are the Monte Carlo simulation results and dashed lines are the numerical solutions from approximate master equations.
}
\label{fig:RR-m}
\end{figure}

\begin{figure}
\centering
\includegraphics[width=.99\linewidth]{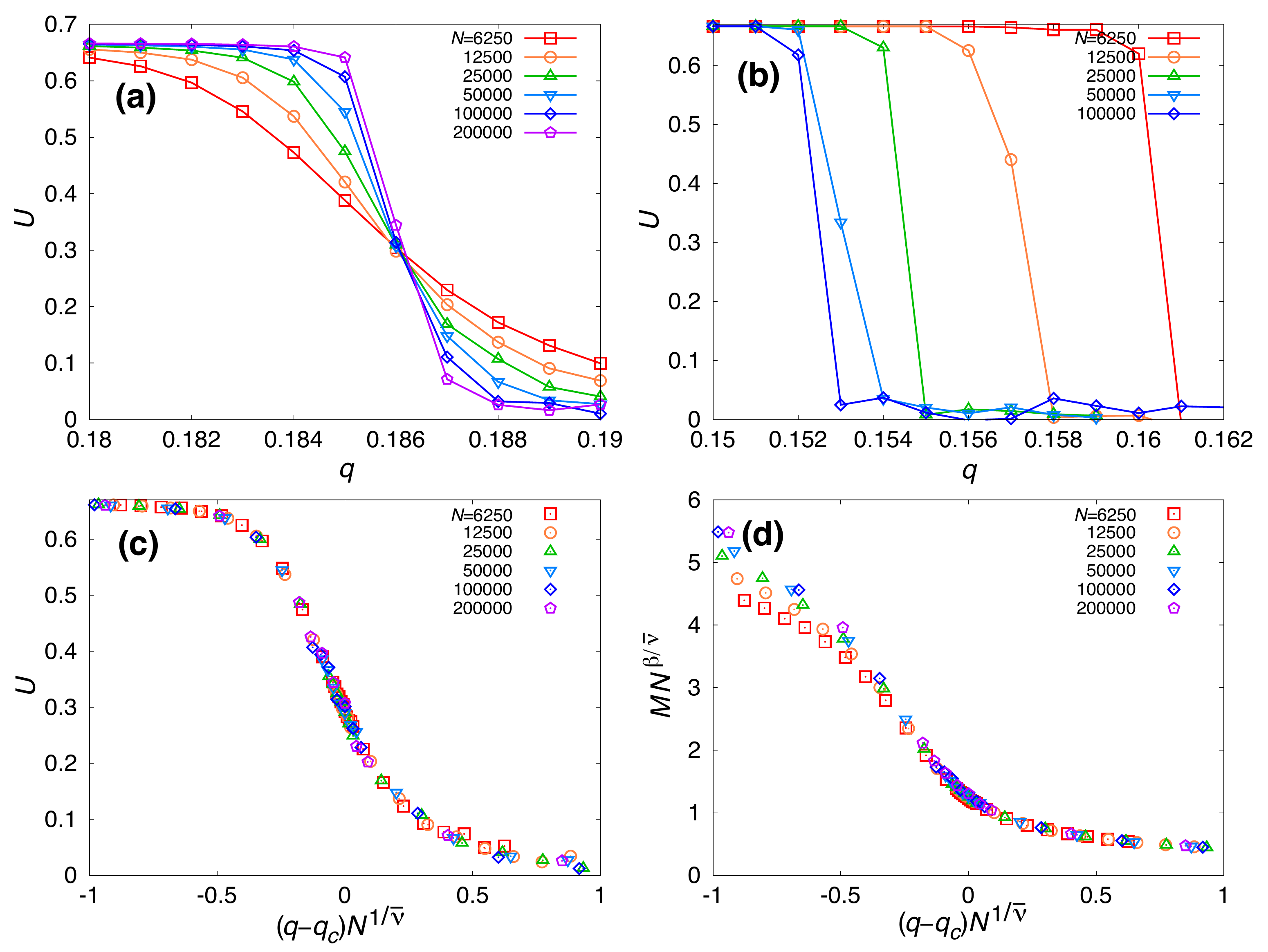}
\caption{
Binder's forth-order cumulant $U$ on $k$-RR duplexes with degree $k=3$ (a) and $k=4$ (b).
Simulation results with network sizes ranging $N=6250$ to $2\times10^5$, averaged over 50 network configurations, are used for data analysis. The transition point is estimated as $q_c=0.1861 \pm 0.0001~(k=3)$ and $q_c= 0.152 \pm 0.001~(k=4)$, respectively. 
Finite-size scaling plots of $U$ and $M$ for $3$-RR duplexes based on Eqs.~(7) and (8) are displayed in (c) and (d), using the mean-field critical exponents $1/\bar{\nu}=1/2$ and $\beta=1/2$. 
}
\label{fig:RR-U}
\end{figure}

\section{Consensus transition on duplexes with random regular network layers} 
We first consider the multiplex models on two-layer (duplex) networks with each layer being random regular network of the same degree $k$ (referred hereafter to as $k$-RR duplexes for short). 
Figure~\ref{fig:RR-m} displays the main results of this study. It highlights two notable features: First, for a given ensemble of networks (that is, given $k$), \textit{AND}-model [Figs.~1(a,b)] tends not only to yield higher consensus level (larger $M$) than \textit{OR}-model [Figs.~1(c,d)], but also to hold the consensus up against higher level of noise (larger $q_c$). 
Secondly, distinct behavior is observed for random regular duplexes with even degrees [Fig.~1(b)] where the consensus transition for \textit{AND}-model turns into a discontinuous one, in marked contrast to the continuous transitions in other cases [Figs.~1(a,c,d)]. In the following, we present further results and discussions to elaborate on these main findings.

\subsection{Nature of consensus transitions}
In order to discern the nature of consensus transitions of the \textit{AND}-model in different cases, we examine the behaviors in the vicinity of the transitions more closely.
To this end, we calculate the Binder's 4th-order cumulant $U$ given by
\begin{eqnarray}
\label{eq:U}
U  =1-\frac{\left\langle \overline{m_t^4} \right\rangle}{3\left\langle \overline{m_t^2} \right\rangle^2} 
\end{eqnarray}
for different network sizes $N$. For $3$-RR duplex, $U(N)$ for different network size $N$ crosses at $q_{\times}=0.1861(1)$ [Fig.~2(a)], which is a signature of continuous transition at $q_c=q_{\times}$. 
On the contrary, for $4$-RR duplex, $U(N)$ does not show clear crossing but the curves tend to converge to a step-function change at $q_{\mathrm{step}}=0.152(1)$ [Fig.~2(b)], which is a typical feature of the discontinuous transition with $q_c=q_{\mathrm{step}}$.

The critical behaviors at the continuous transition of $3$-RR duplex are further examined by the finite-size scaling analyses of $U$ and $M$, using the standard finite-size scaling ansatz,
\begin{eqnarray}
U(N)&=\tilde{U} \left[(q-q_c)N^{1/\bar{\nu}} \right]\\
M(N)&=N^{-\beta/\bar{\nu}} \tilde{M}\left[ (q-q_c)N^{1/\bar{\nu}}\right]~,
\end{eqnarray}
with the order parameter exponent $\beta$ and the correlation volume exponent $\bar{\nu}$. Despite the apparently more abrupt change of $M$ near the transition point of the \textit{AND}-model with odd-$k$ [Fig.~1(a)] than those of the \textit{OR}-model [Figs.~1(c,d)], the finite-size scaling analyses on $3$-RR duplex [Figs.~2(c,d)] suggest that the critical behaviors of the \textit{AND}-model on (\textrm{odd} $k$)-RR duplexes are consistent with the mean-field exponents $\beta=1/2$ and $\bar{\nu}=2$ for the consensus transitions of the majority-vote model in the single-layer networks \cite{Pereira2005}. We also checked that the \textit{OR}-model shows the mean-field behaviors as expected from its spin-flip probability, Eq.~(3).

\begin{figure}
\centering
\includegraphics[width=.99\linewidth]{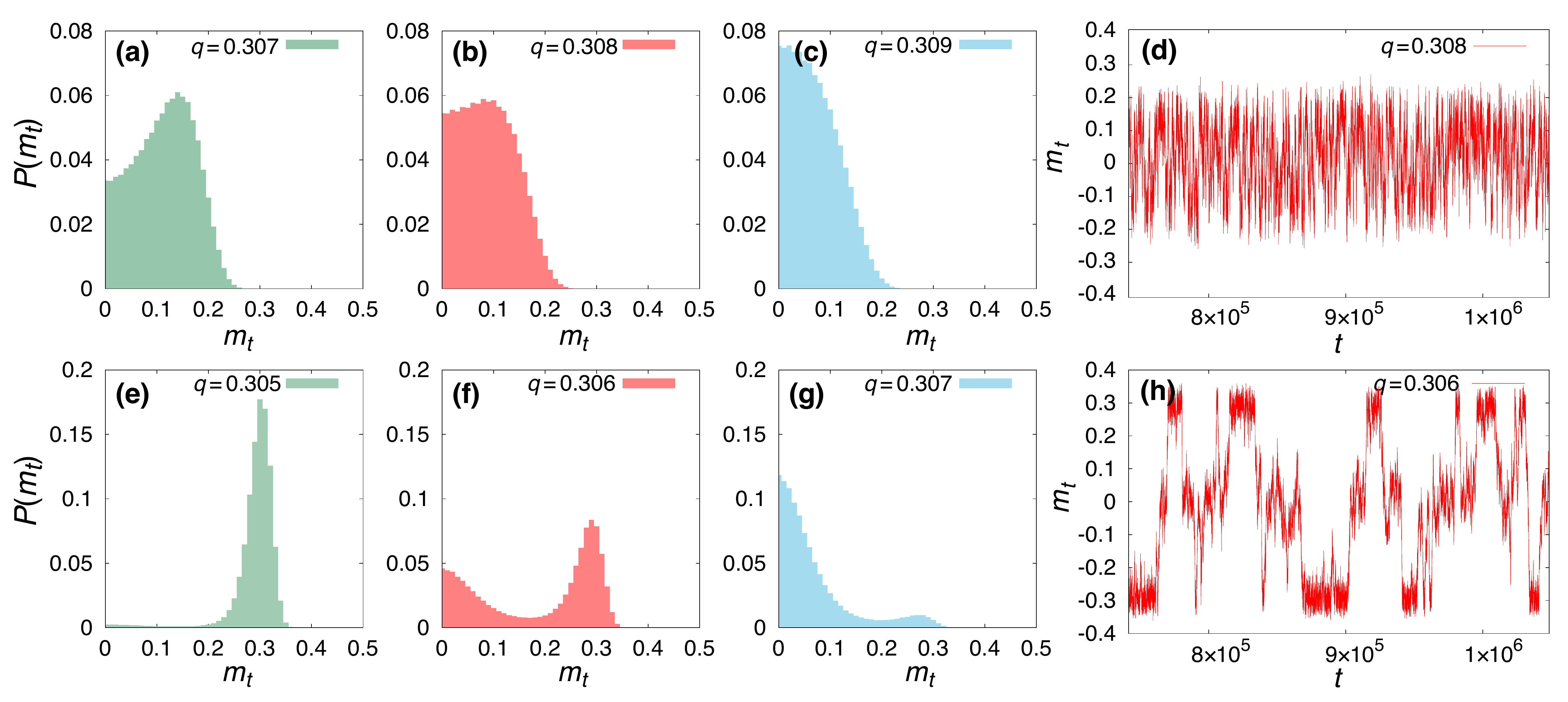}
\caption{
Distributions of instantaneous magnetization $P(m_t)$ (a--c, e--g) and typical transient time trajectory of $m_t$ (d, h) of the \textit{AND}-model on RR duplexes with $k=7$ (a--d) and $k=8$ (e--h).  
Simulations are performed with network size $N=12500$ and the distributions are obtained over the Monte Carlo time interval $[2^{19},2^{20}]$ and over $50$ network configurations. Noise parameters used are $q =\{0.307, 0.308, 0.309\}$ for $k=7$ and $q=\{0.305, 0.306, 0.307\}$ for $k=8$. 
}
\label{fig:histk}
\end{figure}

The discontinuous consensus transitions observed for the \textit{AND}-model on (even $k$)-RR networks deserve closer examinations as well.
To this end, we calculated the probability distribution of instantaneous magnetization at random moment, $P(m_t)$, in the long time limit, which are shown in Fig.~3 for cases of $k=7$ and $k=8$, representative of continuous and discontinuous transition, respectively. The calculations are done for the networks of relatively small size $N=12500$ to observe transient behaviors clearly. The statistics were collected for time interval from $t=2^{19}$ to $t=2^{20}$ Monte Carlo times and averaged over $50$ different networks. For $7$-RR networks, the distribution exhibits a single peak, whose location varies continuously from $m^{\mathrm{peak}}=0$ for $q>q_c$ through $m^{\mathrm{peak}}>0$ for $q<q_c$ [Figs.~3(a--c)]. In contrast, for $8$-RR networks, the distribution develops double peaks as the system approaches to the transition point, across which the location of the higher dominant peak changes discontinuously [Figs.~3(e--g)]. 
Also shown are the typical time series of $m_t$ at the transition point $q_c$ [Figs.~3(d,h)]. The time series fluctuates around $m=0$ for a $7$-RR network [Fig.~3(d)]. On the other hand, it displays transient behavior manifesting stochastic switchings between $m=0$ and $m=\pm m^{\mathrm{peak}}\approx\pm0.3$ [Fig.~3(h)], resulting in the double-peak distribution in Fig.~3(f). These observations support the conclusion that the consensus transition of the \textit{AND}-model on (even $k$)-RR networks is genuinely discontinuous. 

These conclusions for the nature of transitions and the critical behaviors drawn from numerical simulations are corroborated by the analytical calculations using the approximate master equations, shown as lines in Fig.~1. The analytic results agree well with the numerical simulation results in the \textit{OR}-model and in the \textit{AND}-model for large $k$. In the \textit{AND}-model for $k\lesssim 6$, the deviations from numerics are noticeable, but even in this case the approximate analytic calculation predict correctly the qualitative features, such as the nature of transitions and the increase of $q_c$ with $k$.

\begin{figure}
\centering
\includegraphics[width=.99\linewidth]{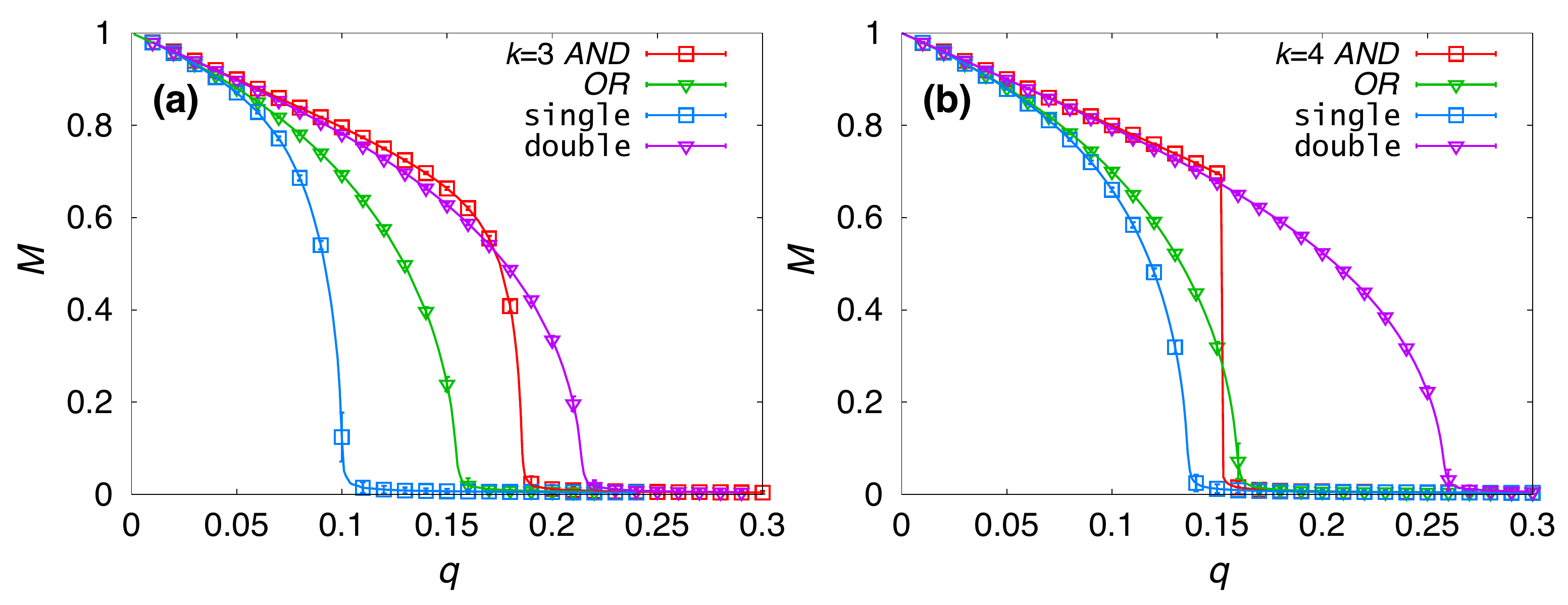}
\caption{The consensus level $M$ on RR-duplexes with degree $k=3$ (a) and $k=4$ (b). The results of \textit{AND}- and \textit{OR}-models are shown along with the results on single-layer network with degree $k$ (\texttt{single}) and aggregated-layer network with degree $2k$ (\texttt{double}). Simulations are done with networks of $N=10^5$ for $t_{\mathrm{final}}=2^{20}$ (\textit{AND}) and $2^{16}$ (others) MC steps, averaged over $50$ different network configurations. Lines are guidelines connecting the numerical simulation results. 
}
\label{fig:compare}
\end{figure}

\subsection{Impact of multiplexity} 
In Fig.~4 we show comparisons between the multiplex models and single-layer majority-vote processes. Along with the multiplex \textit{AND}- and \textit{OR}-model on $k$-RR duplexes, shown are the results of majority-vote model on single-layer RR network with degree $k$ (denoted \texttt{single}) and $2k$ (denoted \texttt{double}), for two examples of $k=3, 4$. We note two features: First, the consensus level of \textit{OR}-model is equal to neither of the \texttt{single} and \texttt{double} cases but lies in the intermediate level between them. Second, for the \textit{AND}-model, the transition point is between those of \texttt{single} and \texttt{double} and the consensus level below $q_c$ is even larger, albeit slightly, than that of \texttt{double}. These results illustrate primary multiplex effects: In understanding the majority-rule-driven consensus dynamics on multiplex social networks, ignoring the presence of other social interaction layer (\texttt{single}) will underestimate both the consensus level ($M$-value) and the tolerance of it against the noise ($q_c$-value); on the other hand, ignoring the multiplexity of interactions and treating them as simple aggregate (\texttt{double}) will generally overestimate the consensus level and its tolerance to noise, but also with the flip-side possibility that even higher consensus could be attained when the conjunctive multiplexity is sufficiently strong as in our \textit{AND}-model. 
It is worth mentioning in passing that for $k=4$, shown in Fig.~4(b), $q_c$ is larger by a small amount for \textit{OR}-model than for \textit{AND}-model in contrast to other cases studied where the latter has significantly higher $q_c$ than the former.

The discontinuous transition displayed by the \textit{AND}-model on (even $k$)-RR duplexes is reminiscent of discontinuous or hybrid transitions in other multiplex models such as the mutual percolation \cite{Son2012} and the multiplex threshold cascade dynamics \cite{Lee2014}. We have assumed for our \textit{AND}-model that a node would not change its state when it does not have a common majority state across all the layers. This creates an ``inertia''-like effect \cite{Chen2017} against the local field produced by the neighboring nodes as well as the stochastic noise parametrized by the $q$ factor. The specific way we formulated the \textit{AND}-model is to maximize this effect. A natural inquiry would follow as to what degree this inertia-like effect could be softened while keeping the transition discontinuous. 
We examined two variations of the \textit{AND}-model for the update rule in the absence of the common majority. The distinctive aspect of the \textit{AND}-model on (even $k$)-RR duplexes compared to that on (odd $k$)-ones is the possibility of the opinion tie in individual layer. Therefore, we focus on the variations on the update rule in the presence of the opinion ties. 
In the first variant, we allow the node to update its opinion at random only if there is no majority (that is, opinion tie) in both layers.  
In the second variant, we allow the node to update its opinion when there is a majority in one layer but no majority (opinion tie) in the other layer. In this case, the node follow the one-layer majority opinion with the probability $1-q$ (and the opposite opinion with the rest probability $q$). 
These two variants implement a less stubborn version of conjunctive majority rule than our original \textit{AND}-model. 
We found both of the variants no longer sustain the discontinuous phase transition displayed by the original \textit{AND}-model on (even $k$)-RR duplexes. 
We also checked that the inertia-like rule of \textit{AND}-model (that is, maintaining its current state when there is no majority) is by itself not strong enough to drive the majority-vote processes on single-layer network towards the discontinuous consensus transition \cite{Chen2017}.

\begin{figure}
\centering
\includegraphics[width=.99\linewidth]{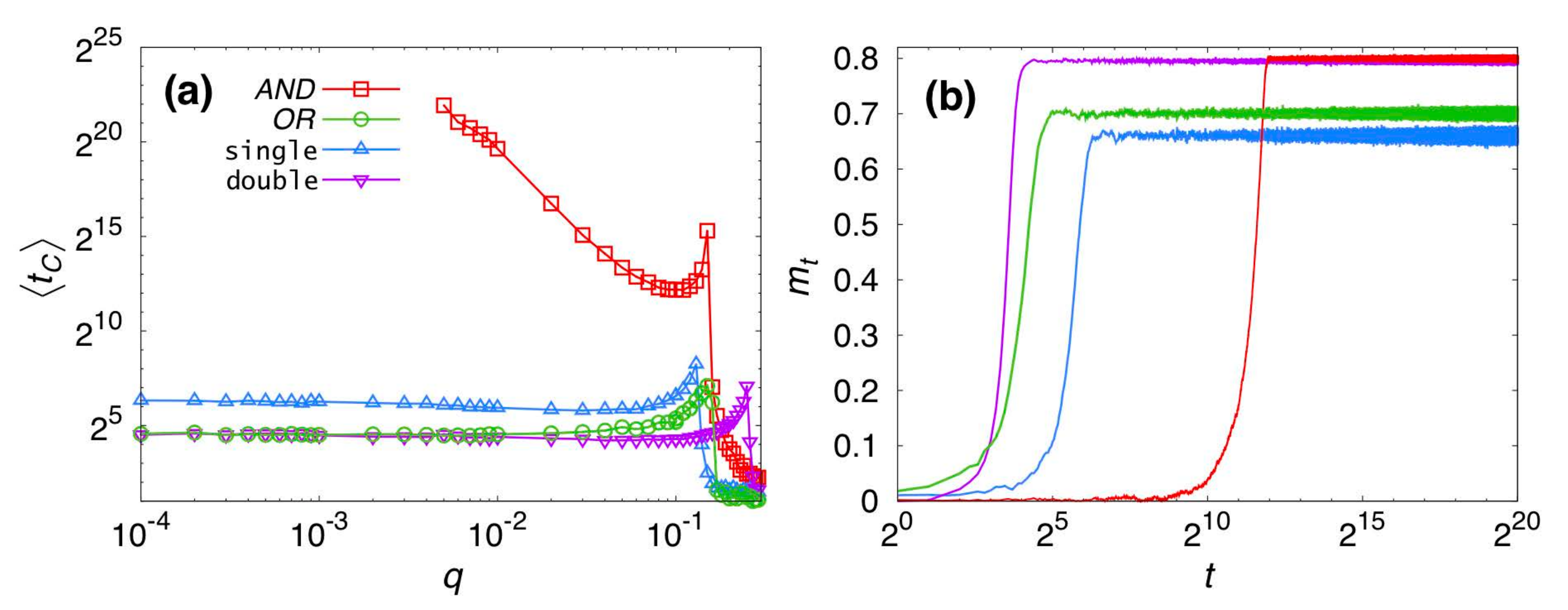}
\caption{The consensus time $\langle t_C\rangle$  (a) and typical time evolution of the instantaneous magnetization $m_t$ at $q=0.1$ (b) on multiplex models on $4$-RR duplexes together with those on \texttt{single} and \texttt{double} networks. Results in (a) is obtained with the networks with size $N=10^{5}$ and averaged over $50$ network configurations.  
}
\label{fig:time}
\end{figure}

\subsection{Time to consensus}
We observed that the \textit{AND}-model could attain larger consensus level than the \textit{OR}-model as well as the single-layer counterparts below $q_c$. 
In achieving such largest consensus level, the inertia-like rule (zero flip-probability for no common majority) is thought to play an important role, by inducing the ratchet-like effect to drive the system towards increasing consensus against the fluctuating local fields and noise. A node without the common majority would ``wait,'' without changing its state, until the common majority is formed in its neighborhood. This could have the effect of slowing down the dynamics towards the consensus. 
To examine the timescale of consensus formation quantitatively, we define the consensus time $t_C$ of each simulation run by 
\begin{eqnarray}
t_C =\inf_{t} \left\{t: \left| m_{t} - \overline{m} \right| < \sigma_m \right\}~.
\end{eqnarray}
where $\overline{m}$ and $\sigma_m$, respectively, are as in Eq.~(5) the time average and standard deviation of $m_t$ at the final time interval ($t_{\mathrm{final}}/2, t_{\mathrm{final}}$), where $t_{\mathrm{final}}$ is the total simulation time (typically we use $t_{\mathrm{final}}=2^{20}$). $t_C$ gives the timescale after which the system can be thought to reach the stationary state. 

Fig.~5 shows the results regarding the consensus time. The mean consensus time $\langle t_C\rangle$ is plotted in Fig.~5(a) and typical time courses of $m_t$ for different models in Fig.~5(b). These results show that the \textit{AND}-model takes considerably longer time to reach the stationary consensus state, often several orders of magnitude longer than other models do. As a result, one might expect that the \textit{AND}-model requires much longer time to arrive the final consensus state than other models (Fig.~5). Furthermore, the evolution of consensus level in time is step-like rather than gradual (in logarithmic timescale). Consequently, if one were concerned with the consensus level attainable by a fixed amount of time, one would arrive at a different conclusion than that for the stationary-state answer in the previous sections. For example, as illustrated in Fig.~5(b), for the $4$-RR duplex with timescale of $t\approx2^7$, the \textit{OR}-model can achieve larger consensus ($t> t_C\approx 2^{5}$) than the \textit{AND}-model, because the \textit{AND}-model requires much longer time ($t<t_C\approx 2^{12}$) to reach its full consensus level and remains nearly zero consensus state at the time of interest $t\approx2^7$. 

The above analyses on time to consensus demonstrate that indeed the slowing-down effect in the \textit{AND}-model can be prominent. Such effect gets more pronounced for weaker noise as $\langle t_C\rangle$ increases considerably as $q$ decreases for the \textit{AND}-model whereas it saturates for other models as seen in Fig.~5(a). It also highlights the importance of consideration of timescales in consensus formation dynamics.

\begin{figure}
\centering
\includegraphics[width=.99\linewidth]{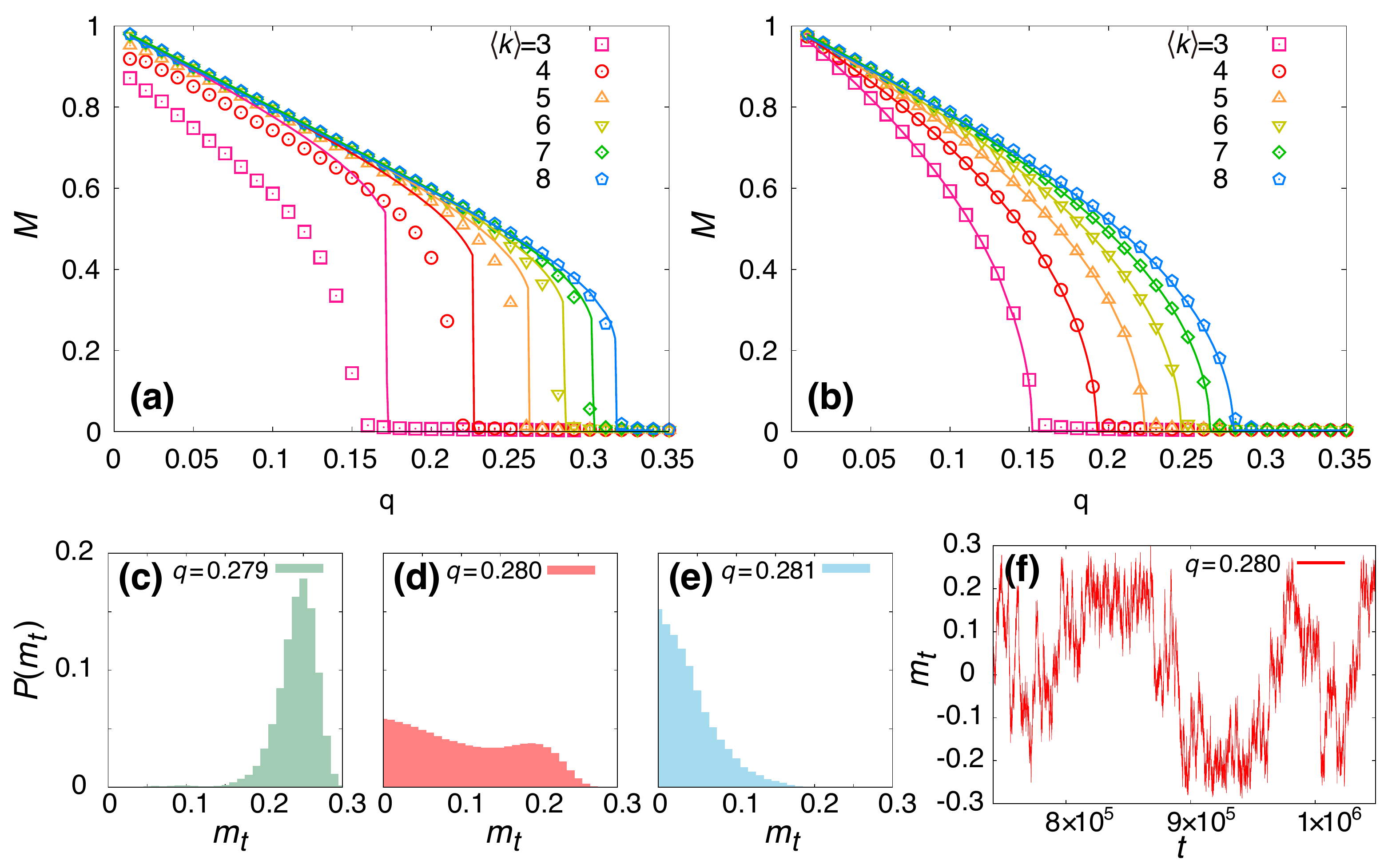}
\caption{(a, b) Consensus level $M$ of \textit{AND}- (a) and \textit{OR}-model (b) on ER duplexes with layer mean-degree $z=3,4,5,6,7,8$. 
Points are Monte Carlo simulation results and lines are the numerical solution of the approximated master equations. 
(c--f) Distributions of instantaneous magnetization $P(m_t)$ (c--e) and typical transient time trajectory of $m_t$ (f) of the \textit{AND}-model in the case of $\langle k\rangle=6$ at $q=\{0.279, 0.280, 0.281\}$, displaying bimodality at $q=0.280$. Simulations are done with networks of  size $N=5\times10^4$ and the distributions are obtained with $50$ network configurations. 
}
\label{fig:ER}
\end{figure}

\section{Majority-vote processes on duplexes with Erd\H{o}s-R{\'e}nyi layers}
The analyses on the RR duplexes in the previous section have unveiled important role of the parity of node degrees in the network on multiplex majority-vote processes, especially the \textit{AND}-model. 
In reality, nodes with different degrees form a network. As a next step,
we consider the \textit{AND}- and \textit{OR} models on duplexes of Erd\H{o}s-R{\'e}nyi network layers of common mean degree $\langle k\rangle$ (hereafter referred to as ER duplexes, for short). 
In ER duplexes some nodes may have no links ($k=0$) in one or all layers.  
The nodes with no links in both layers do not participate to the dynamics. 
For the nodes with neighbors in only one layer, we assume that they work as single-layer network voters, that is, they update their state following Eq.~(1). 

Results of numerical simulations of the \textit{AND}- and \textit{OR}-model on ER duplexes are summarized in Fig.~6. The \textit{AND}-model displays discontinuous transition with nonzero jump of consensus level at the transition point [Fig.~6(a)], whereas the transition is continuous for \textit{OR}-model [Fig.~6(b)].  
The discontinuity in \textit{AND}-model on ER duplexes is weaker than that on (even $k$)-RR duplexes. For example, for ER duplexes with $\langle k\rangle =6$, we observe the bimodal-peaked distribution of magnetization near the transition point and the multistable transient switching dynamics, both indications of discontinuous transition, albeit with less pronounced peaks and noisier switchings  [Figs.~6(c--e)]. This is likely due to the competition between the even-degree nodes that cause discontinuity and the odd-degree nodes that do not. 
The discontinuity gets lessened as the mean degree $\langle k\rangle$ decreases and it becomes barely identifiable for $\langle k\rangle\lesssim 3$. It is not fully clear whether the discontinuity disappears completely or stays at very small nonzero value in this case. More detailed analysis is called for to answer it conclusively. Another notable feature is the growing deviations of the approximate master equation calculation results from the numerical simulation results of the \textit{AND}-model for small $\langle k\rangle$. The calculation results agree with the simulation results very well for the \textit{OR} model and reasonably well for the \textit{AND} on large $\langle k\rangle \gtrsim 6$. On small $\langle k\rangle\lesssim 5$, however, the deviation is appreciable, to the extent that the effectiveness of the approximate master equation formalism is questionable in this case. 

Overall, the qualitative behaviors of the multiplex majority-vote processes on ER duplexes can be understood from their behaviors on RR duplexes. 
These include the nature of transitions (discontinuous for \textit{AND}- and continuous for \textit{OR}-model) and the strength of consensus (larger $M$ and larger $q_c$ of \textit{AND}- than \textit{OR}-model in most mean degrees). Quantitative questions remain as to both the accurate consensus level and the precise nature of transition for the \textit{AND}-model on small mean degrees, which we leave for future study.

\section{Summary and discussion}
In this paper, we have studied majority-vote processes on multiplex networks, by introducing two toy models, the \textit{AND}- and \textit{OR}-model, implementing the layerwise majority-rule in conjunctive and disjunctive manner, respectively. 
The results illustrate the impact of multiplexity. 
Both the multiplex majority-vote processes are found to behave differently from those on the isolated single-layer network and on the network with a simply-aggregated layer, strengthening the premise that the multiplex dynamic processes cannot be reduced to a single-layer one \cite{Min2016,Gomez2013,Brummitt12}.
The \textit{AND}-model can generally attain larger consensus level than \textit{OR}-model but it may take considerably longer time to achieve such a higher consensus, making the available timescale of the process as an important factor as the dynamic rule itself. 
We also found that the \textit{AND}-model can be implemented to induce discontinuous consensus transition on multiplex networks dominated by even-degree nodes, suggesting that the conjunctive layerwise decision rule can be a natural way for the inertia-like factor in the majority-vote processes to be implemented to drive discontinuous phase transitions in multiplex social networks. 

Our study adds to the continuing effort of discerning general effects of multiplexity on dynamic processes on multiplex networks, compared to conventional single-layer ones. In this respect, the majority-vote processes considered in this paper has particular feature that nodes are described by a unique binary state across all the layers and the state evolution does not possess the so-called permanently active property~\cite{Gleeson2007}.
Our results show that discontinuous transition can occur, under right conditions, in this class of dynamic processes. Several questions remain from the perspective of both the current models and the multiplex dynamic processes in general. 
For the current models, issues of immediate interest would include, but not limited to, identifying the weakest condition for the discontinuity in the conjunctive rule of \textit{AND}-type model and improving the analytic approximation method towards better solutions for the \textit{AND}-model. 
Extensions of the present study to more realistic network structures as well as more realistic and detailed layer-wise majority-rules would also be worthwhile. From a broader theoretical perspective, still missing is a generic theoretical framework of multiplex processes, both equilibrium and nonequilibrium, from which one can understand at the fundamental level the governing physics underlying the emergent characteristics such as the discontinuity and/or hysteresis in multiplex systems. We hope this work could contribute to invigorate ongoing endeavors in all these diverse directions.

\section*{Acknowledgments}
This work was supported in part by the National Research Foundation of Korea (NRF) grants funded by the Korea government (MSIT) (No.\ 2017R1A2B2003121). 

\section*{Appendix: Approximate master equations and pair approximation}
Numerical solutions for the majority-vote processes, the \textit{AND}- and \textit{OR}-model, on multiplex networks are obtained by using the approximated master equation formalism \cite{Gleeson2013}.
The binary opinion (spin) states are referred to as \textit{pros} and \textit{cons}. 

\begin{table}
\centering
\begin{tabular}{c|c|c}
\hline
$F_{k,k',m,m'}$ & $(m,m')$ & $R_{k,k',m,m'}$ \\ \hline
$q$ &  $\frac{k}{2}-m>0$ and $\frac{k'}{2}-m'>0$ & $1-q$ \\ \hline
$1-q$ & $\frac{k}{2}-m<0$ and $\frac{k'}{2}-m'<0$ & $q$ \\ \hline
$0$ & otherwise & $0$ \\ \hline
\end{tabular}
\caption{Transition rates for the \textit{AND}-model. Tabulated are the transition rates of turning from pro to con ($F_{k,k',m,m'}$) and those of turning from con to pro ($R_{k,k',m,m'}$) when a node has $(m,m')$ con-neighbors among its $(k,k')$-degree.}
\label{tb:AND}
\end{table}
\begin{table}
\centering
\begin{tabular}{c|c|c}
\hline
$F_{k,k',m,m'}$ & $(m,m')$ & $R_{k,k',m,m'}$\\ \hline
$q$  & $\frac{k}{2}-m>0$ and $\frac{k'}{2}-m'>0$ & $1-q$ \\ \hline
$1-q$ & $\frac{k}{2}-m<0$ and $\frac{k'}{2}-m'<0$ & $q$ \\ \hline
\multirow{2}{*}{$\frac{1}{2}$} & $(\frac{k}{2}-m)(\frac{k'}{2}-m')<0$ & \multirow{2}{*}{$\frac{1}{2}$} \\ 
& $\frac{k}{2}-m=\frac{k'}{2}-m'=0$ &  \\ \hline
\multirow{2}{*}{$\frac{1}{4}+\frac{q}{2}$} & $\frac{k}{2}-m=0$ and $\frac{k'}{2}-m'>0$ & \multirow{2}{*}{$\frac{3}{4}-\frac{q}{2}$} \\ 
& $\frac{k}{2}-m>0$ and $\frac{k'}{2}-m'=0$ & \\ \hline
\multirow{2}{*}{$\frac{3}{4}-\frac{q}{2}$} & $\frac{k}{2}-m<0$ and $\frac{k'}{2}-m'=0$ & \multirow{2}{*}{$\frac{1}{4}+\frac{q}{2}$} \\ 
& $\frac{k}{2}-m=0$ and $\frac{k'}{2}-m'<0$ &  \\ \hline
\end{tabular}
\caption{Transition rates for the \textit{OR}-model. 
}
\label{tb:OR}
\end{table}

The master equation is set up by following Ref.~\cite{Gleeson2013} with modifications to account for the multiplexity. The multiplex network is characterized by the joint degree distribution $P_{k,k'}$ with $k$ denoting the degree in the first layer and $k'$ that in the second. We define $p_{k,k',m,m'}$ to be the fraction of pros among the nodes with $(k,k')$-degree (denoting degree $k$ in the first layer and $k'$ in the second) and $(m,m')$ con-neighbors (denoting $m$ neighbors with con-opinion in the first layer and $m'$ in the second). 
Likewise, $c_{k,k',m,m'}$ is the fraction of cons among the nodes with $(k,k')$-degree and $(m,m')$ con-neighbors.
The layer-wise majority-vote rule is translated into the transition rates $F_{k,k',m,m'}$ and $R_{k,k',m,m'}$:
$F_{k,k',m,m'}$ denotes the probability that a pro-opinion node becomes a con-opinion when it has $(k,k')$-degree and $(m,m')$-con-neighbors.  
Similarly the reverse transition rate $R_{k,k',m,m'}$ is the probability that a con-opinion node turns to a pro-opinion one.
Using these rates, the master equation for the pros and the cons, respectively, is written as
\begin{eqnarray}
\label{eq:meq-pors}
\frac{d}{dt}p_{k,k',m,m'}=&-F_{k,k',m,m'}p_{k,k',m,m'}+R_{k,k',m,m'}c_{k,k',m,m'} \nonumber\\
&-\beta^p(k-m)p_{k,k',m,m'}+\beta^p(k-m+1)p_{k,k',m-1,m'} \nonumber\\
&-\gamma^p~m~p_{k,k',m,m'}+\gamma^p (m+1)p_{k,k',m+1,m'} \nonumber\\
&-\beta^p(k'-m')p_{k,k',m,m'}+\beta^p(k'-m'+1)p_{k,k',m,m'-1} \nonumber\\
&-\gamma^p m' p_{k,k',m,m'}+\gamma^c (m'+1)p_{k,k',m,m'+1}~,
\end{eqnarray}
and
\begin{eqnarray}
\label{eq:meq-cons}
\frac{d}{dt}c_{k,k',m,m'}=&-R_{k,k',m,m'}p_{k,k',m,m'}+F_{k,k',m,m'}c_{k,k',m,m'} \nonumber\\
&-\beta^c(k-m)c_{k,k',m,m'}+\beta^c(k-m+1)c_{k,k',m-1,m'} \nonumber\\
&-\gamma^c m c_{k,k',m,m'}+\gamma^c (m+1)c_{k,k',m+1,m'} \nonumber\\
&-\beta^c(k'-m')c_{k,k',m,m'}+\beta^c(k'-m'+1)c_{k,k',m,m'-1} \nonumber\\
&-\gamma^c m' p_{k,k',m,m'}+\gamma^c (m'+1)c_{k,k',m,m'+1}~,
\end{eqnarray}
where the coefficients $F$'s and $R$'s are given in Table~1 and 2 for the \textit{AND}- and \textit{OR}-model, respectively.
$\beta$ and $\gamma$ factors denote the following: $\beta^p$ is the probability that a pro-pro edge changes to a con-pro edge and $\gamma^p$ is the probability that a con-pro edge becomes a pro-pro edge.
Likewise, $\beta^c$ is the probability that a pro-con edge changes to a con-con edge and $\gamma^C$ is the probability that a con-con edge becomes a pro-con edge. 
Under pair approximation, these factors are given by
\begin{eqnarray}
\beta^{p}&=\frac{\left\langle \sum_{m,m'=0}^{k,k'} (k-m+k'-m')F_{k,k',m,m'}p_{k,k',m,m'} \right\rangle}
{\left\langle \sum_{m,m'=0}^{k,k'} (k-m+k'-m')p_{k,k',m,m'} \right\rangle}~,\nonumber\\ 
\gamma^{p}&=\frac{\left\langle \sum_{m,m'=0}^{k,k'} (k-m+k'-m')R_{k,k',m,m'}c_{k,k',m,m'} \right\rangle}
{\left\langle \sum_{m,m'=0}^{k,k'} (k-m+k'-m')c_{k,k',m,m'} \right\rangle}~,\nonumber\\
\beta^{c}&=\frac{\left\langle \sum_{m, m'=0}^{k,k'} (m+m')F_{k,k',m,m'}p_{k,k',m,m'} \right\rangle}
{\left\langle \sum_{m,m'=0}^{k,k'} (m+m') p_{k,k',m,m'} \right\rangle}~,\nonumber\\
\gamma^{c}&=\frac{\left\langle \sum_{m,m'=0}^{k,k'} (m+m') R_{k,k',m,m'}c_{k,k',m,m'} \right\rangle}
{\left\langle \sum_{m,m'=0}^{k,k'} (m+m') c_{k,k',m,m'} \right\rangle}~,
\end{eqnarray}
Here the average $\langle \cdots\rangle$ is taken over the joint degree distribution $P_{k,k'}$. Equations (10) and (11) are iterated until stationary values of $p_{k,k',m,m'}$ and $c_{k,k',m,m'}$ are obtained. 

Finally one can calculate the fraction of cons of degree $(k,k')$ as 
\begin{eqnarray}
\rho_{k,k'}=1-\sum_m^k\sum_{m'}^{k'}p_{k,k',m,m'}, 
\end{eqnarray}
from which the consensus level $M$ is obtained as 
\begin{eqnarray}
M=\left|1-2\sum_{k,k'}\rho_{k,k'}P_{k,k'}\right|~,
\end{eqnarray}
which is plotted in Figs.~1, 4, and 6.

\section*{References}
\bibliography{voter}
\bibliographystyle{ieeetr}

\end{document}